\documentclass[12pt]{article}
\usepackage{a4wide}
\usepackage{amssymb}
\usepackage{amsmath}
\usepackage{graphicx}
\usepackage{slashed}
\usepackage{verbatim}

\def\XXint#1#2#3{{\setbox0=\hbox{$#1{#2#3}{\int}$}
     \vcenter{\hbox{$#2#3$}}\kern-.52\wd0}}

\newcommand{\be}{\begin{equation}}\newcommand{\ee}{\end{equation}}
\newcommand{\bea}{\begin{eqnarray}} \newcommand{\eea}{\end{eqnarray}}

\def\p{\partial}

\makeatletter \@addtoreset{equation}{section} \makeatother

\begin{document}

\setcounter{table}{0}

\begin{flushright}\footnotesize

\texttt{ICCUB-17-010}
\vspace{0.6cm}
\end{flushright}

\mbox{}
\vspace{0truecm}
\linespread{1.1}

\centerline{\LARGE \bf ${\cal N}= 2$ Chern-Simons-Matter Theories}
\bigskip
\centerline{\LARGE \bf   Without Vortices }

\vspace{.5cm}

 \centerline{\LARGE \bf }

\vspace{1.5truecm}

\centerline{   {\large \bf Jorge G. Russo ${}^{a,b}$} \footnote{jorge.russo@icrea.cat}
   {\bf and}    {\large \bf  Fidel A. Schaposnik${}^{c}$} \footnote{fidel@fisica.unlp.edu.ar}}

\vspace{1cm}
\centerline{{\it  ${}^a$ Instituci\'o Catalana de Recerca i Estudis Avan\c{c}ats (ICREA)}} \centerline{{\it Pg.Lluis Compayns, 23, 08010 Barcelona, Spain}}
\medskip
\centerline{{\it  ${}^b$  Departament de F\'\i sica Cu\' antica i Astrof\'\i sica and Institut de Ci\`encies del Cosmos}} \centerline{{\it Universitat de Barcelona, Mart\'i Franqu\`es, 1, 08028
Barcelona, Spain }}

\medskip
\centerline{{\it  ${}^c$ Departamento de F\'\i sica-UNLP/IFLP-CONICET}} 
\centerline{{\it CC 67, 1900 La Plata, Argentina}}
\medskip

\vspace{1cm}

\centerline{\bf ABSTRACT}
\medskip

We study ${\cal N}=2$ Chern-Simons-matter theories with gauge group  $U_{k_1}(1)\times U_{k_2}(1)$. We find that, when 
$k_1+k_2=0$, the partition function computed by localization dramatically simplifies and collapses to a single term. 
We show that the same condition  prevents the theory from having supersymmetric vortex configurations. 
The theories include mass-deformed ABJM theory with $U(1)_{k}\times U_{-k}(1)$  gauge group as a particular case. 
Similar features are shared by a class of CS-matter theories with 
gauge group $U_{k_1}(1)\times \cdots \times U_{k_N}(1)$.

\noindent

\newpage

\section{Introduction}

In the last few years, the use of the localization principle in three-dimensional supersymmetric gauge theories  \cite{Kapustin:2009kz,Kapustin:2010xq,Jafferis:2010un,Hama:2010av} led to a number of remarkable results which unveiled the  role of non-perturbative effects in three dimensions and the precise way they
contribute to supersymmetric observables.
The expected non-perturbative contributions in three-dimensional supersymmetric gauge theories come from vortex and antivortex configurations.
In particular,  the partition function on the squashed sphere ${\mathbb S}^3_b$, computed in \cite{Hama:2011ea},
can be expressed as infinite sums where each term contains the product of the vortex times the antivortex partition function \cite{Pasquetti:2011fj}.
This structure  also appeared in ${\cal N}=2$ superconformal indices \cite{Dimofte:2011py,Hwang:2012jh} and general properties underlying this decomposition in ``holomorphic blocks" were further explored in \cite{Beem:2012mb,Chen:2013pha,Yoshida:2014ssa}.

The physical origin of the non-perturbative terms, and its connection to vortices, can be understood in a more direct way if one implements the method of localization by adding a different deformation term to the path integral, in such a way that the classical supersymmetric configurations contributing
to the partition function are precisely vortices at the north pole and antivortices at the south pole of ${\mathbb S}^3_b$ \cite{Fujitsuka:2013fga,
Benini:2013yva}. This alternative localization, called ``Higgs branch localization", was first discovered in \cite{Benini:2012ui,Doroud:2012xw} in the context of two-dimensional ${\cal N}=(2,2)$ 
gauge theories on ${\mathbb S}^2$. 
Many other remarkable phenomena appeared in related works,  in particular, mirror duality exchanging vortex loop  and Wilson loop operators \cite{Kapustin:2012iw,Drukker:2012sr} (see  \cite{Willett:2016adv} for a recent review and a more complete list of references).

Supersymmetric localization reduces the problem of computing a highly complicated functional integral to a
far much simpler finite-dimensional integral.
The exact partition function in  the different ${\cal N}=2$ theories has, nonetheless,  an extremely rich and complicated structure, encapsulating
interesting gauge-theory phenomena in an exact formula. The integrals can be computed by residues, leading to long expressions
representing the sum over vortex and antivortex partition functions described above.
A natural question is whether there are cases where this extremely complicated structure simplifies.
In this note we identify one example where a huge simplification occurs and disclose the physical origin of such simplification.
We consider  three-dimensional 
${\cal N}=2$ supersymmetric Chern-Simons-matter gauge theories on the squashed sphere ${\mathbb S}^3_b$ with gauge group $U_{k_1}(1)\times U_{k_2}(1)$, with matter charged under both gauge groups.
We will find a peculiar phenomenon. For generic parameters, the theory contains vortices and antivortices associated with north and south poles of 
${\mathbb S}^3_b$, with the expected partition function
factorizing in terms of holomorphic blocks. However, when the couplings satisfy a certain condition, supersymmetric vortex configurations are
no longer possible: the theory then contains a unique, topologically trivial vacuum and the partition function reduces to a single term
(yet with highly non-trivial dependence on the couplings). We will also discuss this phenomenon in terms of the effective potential.

The paper is organized as follows. In section 2 we give a brief review of three-dimensional ${\cal N}=2$ gauge theories on ${\mathbb S}^3_b$,
with a focus on   theories with $U(1)\times U(1)$ gauge group.
In section 3 we consider  $U(1)_k\times U(1)_{-k}$ ABJM theory with mass and Fayet-Iliopoulos (FI) deformations and compute the partition function on the three-sphere. 
We show that the integrals can be carried out in a straightforward way, leading to a very simple compact formula for the
partition function. 
In section 4 we consider a  $U(1)\times U(1)$ gauge theory with arbitrary Chern-Simons levels $k_1,\ k_2$ and more
 general matter content.
We show that a similar simplification takes place, both on the three-sphere ${\mathbb S}^3$ and on the ellipsoid
${\mathbb S}^3_b$, provided the parameters satisfy a 
certain constraint. Finally, in section 5, we study supersymmetric vortex configurations in flat space for the general model of section 4 and show
that all vortices disappear when the same condition on the parameters is imposed.
We also show that, for any arbitrary parameters not satisfying this condition, the theory has vortices with an action compatible with
the vortex counting parameter that one derives from the partition function on the ellipsoid.

\section{${\cal N}=2$ supersymmetric gauge theories on the ellipsoid}

We consider the three-ellipsoid with $U(1)\times U(1)$ isometry as in \cite{Hama:2011ea}. The three-ellipsoid is defined by the hypersurface
\be
x_0^2+x_1^2+x_2^2+x_3^2=1\ ,
\ee
with metric
\be
ds^2 = \ell^2 (dx_0^2+dx_1^2) +\tilde \ell^2 (dx_2^2+dx_3^2) \ .
\ee
Introducing coordinates
\be
(x_0,x_1,x_2,x_3)= (\cos\theta \cos\varphi_2 ,\cos\theta \sin\varphi_2,\sin\theta \cos\varphi_1,\sin\theta \sin\varphi_1)\ ,
\ee
the metric takes the form
\be
ds^2 =r^2 \left( f(\theta)^2 d\theta^2 +b^2\sin^2\theta d\varphi_1^2 + b^{-2}\cos^2\theta d\varphi_2^2 \right)\ ,
\label{un}
\ee
$$
b\equiv\sqrt{\tilde \ell /\ell} \ ,\qquad r\equiv \sqrt{\ell \tilde \ell}\ ,\qquad  f(\theta ) \equiv
\sqrt{b^{-2}\sin^2\theta +b^2\cos^2\theta }\ .
$$

Here we shall study ${\cal N}=2$ supersymmetric gauge theories on this space, with gauge group
$U(1)\times U(1)$ and chiral matter. The theories thus have two vector multiplets $(A_1,\sigma_1,\lambda_1,\bar \lambda_1, D_1)$ and $(A_2,\sigma_2,\lambda_2,\bar \lambda_2, D_2)$.

The three-dimensional action  contains  Chern-Simons terms for each $U(1)$ gauge group, i.e.
\be
\label{csact}
 S_{\rm CS}[k] =  i \frac{k}{4\pi} \int A\wedge dA- i \frac{k}{4\pi} \int d^3 x \sqrt{g} (-\bar \lambda \lambda+2 D  \sigma )\ ,
\ee
with general Chern-Simons levels $k_1,\ k_2$.

The FI deformations can be constructed as usual by coupling the vector multiplets to  ${\cal N}=2$ background vector multiplets
$( (\tilde A_a)_\mu,\tilde \sigma_a,\tilde \lambda_a, \bar{\tilde {\lambda}}_a ,\tilde D_a)$, $a=1,2$.
One gets
\be
S_{\rm FI}= \frac{i}{4\pi} \int d^3 x \sqrt{g} (\tilde D_a  \sigma_a +\tilde \sigma_a D_a) ,\qquad a=1,2\ .
\ee

Chiral matter may couple  to both vector multiplets $V_1= (A_1,\sigma_1,\lambda_1,\bar \lambda_1, D_1)$, $V_2=(A_2,\sigma_2,\lambda_2,\bar \lambda_2, D_2)$,
with some given charges $q_1,\ q_2$, so that the covariant derivative is $D_\mu \phi =\p_\mu \phi -i q_1 (A_1)_\mu \phi-i q_2 (A_2)_\mu \phi$.
Defining a vector multiplet $\hat V=q_1 V_1+q_2 V_2$, with components $\hat V= (\hat A,\hat \sigma,\hat \lambda,\bar{\hat \lambda},\hat  D)$,
the action for a chiral multiplet of R-charge $ \Delta $ is then given by (we follow the conventions of \cite{Fujitsuka:2013fga})
\bea
S_{\rm matter} &=& \int d^3 x \sqrt{g} \bigg( D_\mu \bar\phi D^\mu \phi -i \bar \psi \gamma^\mu
    D_\mu \psi +\frac{ \Delta (2-\Delta )}{r^2f^2}\bar\phi \phi + i \bar \phi \hat D \phi-\frac{2\Delta -1}{2rf}\bar\psi\psi
\nonumber\\
    &+& 
    i\bar\psi \hat \sigma \psi +i\bar\psi \hat \lambda \phi -i \bar\phi \bar{\hat\lambda} \psi +\bar \phi {\hat\sigma}^2\phi 
+\frac{i(2\Delta -1)}{rf}\bar \phi \hat \sigma \phi+\bar F F\bigg)\ ,
\label{mmmm}
\eea
where 
$\gamma^\mu $ are the Pauli matrices. An ${\cal N}=2$ preserving mass deformation can be added in the usual way  by coupling the chiral fields to vector multiplets associated with the flavor symmetry. Real masses $m_i$ then correspond to the expectation values of the scalar fields of these background vector multiplets. 

We will consider models with $N_f$  chiral multiplets having   the same charges $q_1,\ q_2$ and $N_a$ chiral multiplets with
charges $-q_1,-q_2$,
with $q_1 q_2\neq 0$ (the case $q_1 q_2=0$ leads to a decoupled  $U(1)$ Chern-Simons-matter theory, which has already been
 studied in the literature, see e.g. \cite{Hong:1990yh,Jackiw:1990aw}.). 
For these models, it is convenient to normalize the gauge fields by setting e.g. $q_1=1,\ q_2=-1$. 
This normalization rescales the CS levels $k_1,\ k_2$ (which in the abelian case on ${\mathbb S}^3_b$ do not need to  be quantized).\footnote{In non-trivial three-dimensional manifolds
 with non-contractible one-cycles  the normalization of the gauge fields must be such that the Chern-Simons levels  $k_1,\ k_2$ are quantized 
for the abelian theory to be invariant under large gauge transformations.}

We will need the supersymmetric transformations for the fermions, which are as follows 
\bea
&&\delta \lambda =\left( \frac{1}2 \epsilon_{\mu\nu\rho}F^{\nu\rho} -\p_\mu \sigma
\right) \gamma^\mu \epsilon - iD\epsilon-\frac{i}{rf}\sigma \epsilon\ ,
\nonumber\\
&&\delta \bar \lambda =\left( \frac{1}2 \epsilon_{\mu\nu\rho}F^{\nu\rho} +\p_\mu \sigma
\right) \gamma^\mu \bar \epsilon + iD\bar \epsilon+\frac{i}{rf}\sigma \bar \epsilon\ ,
\label{ate}
\eea
and
 \bea
 &&\delta \psi = -\gamma^\mu \epsilon D_\mu \phi -\epsilon \hat\sigma \phi - \frac{i\Delta}{rf}\epsilon\phi  +i\bar \epsilon F\ ,
 \nonumber\\
&& \delta \bar \psi = -\gamma^\mu \bar \epsilon D_\mu \bar \phi -\bar \epsilon \hat \sigma \bar \phi- \frac{i\Delta}{rf}
\bar\epsilon\bar \phi  +i \epsilon  \bar F\ .
\label{bate}
\eea

Introducing the localizing term for Coulomb branch localization as in \cite{Hama:2011ea},
the fields  localize to the configuration
\be
D_1=- \frac{\sigma_1}{rf}\ ,\qquad D_2 = - \frac{\sigma_2}{rf}\ , 
\ee
with constant $\sigma_1,\ \sigma_2$, other fields localizing to vanishing values. 
Similarly, supersymmetry requires that the background fields appearing in the FI deformations also satisfy
$\hat D_a=- \frac{\hat \sigma_a}{rf}$. Integrating over $\theta $,
the FI terms localize to $2\pi i \eta_a \sigma_a $, where $\eta_a $, $ a=1,2$ represent constant parameters related to the values of the background fields.

For the present theory, using the rules derived  \cite{Hama:2011ea} (generalizing the formula for the partition function on the three-sphere \cite{Kapustin:2009kz}),
 the exact partition function has the form
\be
Z=\int d\sigma_1 d\sigma_2 \ e^{-i\pi k_1  \sigma_1^2 -i\pi k_2 \sigma_2^2
+2\pi i(\eta_1 \sigma_1+\eta_2\sigma_2) }\  Z^{\rm chiral}_{{\rm 1-loop}} (\sigma_1,\, \sigma_2 )\ ,
\ee
where $Z^{\rm chiral}_{{\rm 1-loop}}$ represents the one-loop determinant coming from the matter sector.

\section{$U(1)\times U(1)$ ABJM theory with FI and mass deformations}

The first model is inspired by ABJM theory \cite{abjm}.
Specifically, the $U(1)\times U(1)$  model contains CS actions with opposite levels. There are  two chiral multiplets with 
$\Delta=1/2$, gauge charges $(1,-1)$  and mass parameters $\pm m$ and two antichiral multiplets with the same masses and
gauge charges  $(-1,1)$.
In addition, we shall also include a   FI term for the diagonal $U(1)$. 
One can anticipate that this theory will be particularly simple, since for the abelian $U(1)\times U(1)$ ABJM theory the sixth-order potential vanishes \cite{abjm},   leaving only the mass deformations and therefore a theory of  two chiral and two antichiral {\it free} superfields.


For the theory on the three-dimensional ellipsoid (\ref{un}), the action of the model is defined by
\be
S =  \left(S_{\rm CS}[k]  +S_{\rm FI}[\eta] \right)_1+\left(S_{\rm CS}[-k] +S_{\rm FI}[\eta]\right)_2 +S_{\rm matter}   \ ,
\ee
where the different terms have been defined above.

We start with the simplest case where $b=1$, corresponding to the sphere limit of the ellipsoid. In this case, $f(\theta)=1$. 
In the next section we will generalize the formulas for a model with  arbitrary $\Delta $ and $b$ parameters.
The partition function is given by 
\be
Z=\int d\sigma_1 d\sigma_2  \ \frac{e^{-i\pi k (\sigma_1^2-\sigma_2^2)+2\pi i\eta (\sigma_1+\sigma_2) }}
{\cosh\big(\pi (\sigma_1-\sigma_2+m)\big)\cosh\big(\pi(\sigma_1-\sigma_2-m)\big)}\ . 
\ee
This is the same expression for the mass/FI deformed ABJM partition function given in \cite{Kapustin:2010xq} particularized to $N=1$.
Now we introduce new integration variables:
\be
\sigma_+ = \frac{\sigma_1+\sigma_2}{2}\ ,\qquad
\sigma_- = \frac{\sigma_1-\sigma_2}{2}\ .
\label{plusminus}
\ee
The partition function becomes
\be
\label{sigg}
Z=2\int d\sigma_+ d\sigma_-  \ \frac{e^{-4\pi ik \sigma_+\sigma_-
+4\pi i\eta \sigma_+ }}{\cosh\big(\pi(2\sigma_- +m)\big)\cosh\big(\pi(2\sigma_- - m)\big)}\ .
\ee
Integrating over $\sigma_+$, we get a Dirac $\delta $-function
\be
Z=  \int  d\sigma_-  \ \frac{  \delta(k\sigma_-- \eta )}{\cosh\big(\pi(2\sigma_- +m)\big)\cosh\big(\pi (2\sigma_- - m)\big)} \ .
\ee
Therefore, the partition function has the compact form
\be
Z= \frac{1}{|k|} \ \frac{ 1}{\cosh \big(\pi (\frac{2\eta}{k} +m)\big)\cosh\big( \pi (\frac{2\eta}{k} - m)\big)}\ .
\ee
Note that the expansion in powers of $1/k$ corresponds to the perturbative expansion.
It has a finite radius of convergence $1/k_0$, determined by the first zero of
 $\cosh\pi (2\eta/k \pm m)$ in the complex $1/k$-plane, i.e.
$$
\frac{1}{k_0} =\left| \frac{1}{2\eta} \left(m \pm \frac{i}{2}\right)\right|\ .
$$ 
This is in contradistinction
with the  behavior of the weak coupling perturbation series in more general ${\cal N}=2$ supersymmetric
gauge theories, which is asymptotic \cite{Russo:2012kj,Aniceto:2014hoa,Honda:2016vmv}.

For integer $k$, in some cases the partition function
on  ${\mathbb S}^3$ has a finite number of terms (see \cite{Kapustin:2010mh,Okuyama:2011su,Awata:2012jb,Russo:2014bda,Russo:2015exa,Giasemidis:2015ial} for many examples).
 This can be illustrated by $U(1)$ ${\cal N}=2$ Chern-Simons theory with a FI deformation, coupled to a pair of massless chiral fields of $\Delta=1/2$ and opposite gauge charges.
The partition function is given by
\be
Z = \int d\sigma \ e^{-i \pi k \sigma^2} \
 \frac{e^{2\pi i\eta\sigma}}{\cosh\big(\pi \sigma\big)} \ .
\label{Zmodelc}
\ee
Integrating by residues, it might seem that we get an infinite sum coming from the poles of the $\cosh\big(\pi \sigma\big)$ on the imaginary axes. 
However, some care is needed in order to choose the integration contour, since
the integrand does not decay exponentially on a large semicircle.
It is convenient to go to the ``dual" representation by writing
\be
\frac{1}{\cosh\big(\pi \sigma\big)} =\int d\tau \frac{e^{2\pi i \tau \sigma}}{\cosh(\pi\tau )}\ .
\ee
Computing the Gaussian integral over $\sigma $, and shifting $\tau\to \tau-\eta$, we find
\be
Z = \frac{e^{-\frac{i\pi}{4}}}{\sqrt{k}}\int d\tau \ e^{ \frac{i \pi \tau^2}{k}} \
 \frac{1}{\cosh\big(\pi (\tau-\eta)\big)} \ .
\label{Zmodelcc}
\ee
This is a Mordell integral \cite{mordell} (see \cite{Russo:2014bda,Giasemidis:2015ial} for explicit examples
in the context of ${\cal N}=2$ CS theories).
For integer $k$, the integral
can be computed  by choosing an appropriate rectangular contour,
leading to a finite sum \cite{Russo:2014bda}
\be
Z=-\frac{2e^{i\pi (x-k/4)}}{e^{2i\pi x}-1}\left( \sqrt{\frac{-i}{k}} \sum_{n=0}^{k-1} e^{-\frac{i\pi}{k} (x-\frac{k}{2}-n)^2}+i e^{2i\pi x}\right)\ ,
\ee
with $x\equiv -i\eta -1/2$. For non-integer $k$, the integral gives rise to an infinite sum which can be  expressed in terms of $\theta $ functions
\cite{mordell}.

On the other hand, on the ellipsoid, the partition function with $k_1+k_2\neq 0$ contains an infinite series of terms
and they represent vortex contributions as in \cite{Pasquetti:2011fj,Fujitsuka:2013fga,Benini:2013yva,Kapustin:2012iw,Drukker:2012sr}. 
We discuss the ellipsoid partition function in the next section. 


\section{More general $U(1)\times U(1)$  model}

In this section we consider a more general model where the Chern-Simons levels for the  $U(1)\times U(1)$  gauge group
are $(k_1,k_2)$, with general matter content.\footnote{Deformations of ABJM theory to general Chern-Simons levels $k_1,\ k_2$ have been proposed
to have  an holographic interpretation  in terms of $AdS_4$ backgrounds with non-zero Romans mass
\cite{Gaiotto:2009mv}.}

\subsubsection*{Partition function on the three-sphere}

We first consider 
$2N_f$  chiral fields with charges $(1,-1)$ and
$2N_f$  chiral fields with  charges $(-1,1)$, all with the same R-charge $\Delta=1/2 $ and masses $\pm m$.
The partition function on the three-sphere is now given by
\be
\label{zeva}
Z=\int d\sigma_1 d\sigma_2  \ \frac{e^{-i\pi k_1 \sigma_1^2 -i\pi k_2 \sigma_2^2
+2\pi i (\eta_1 \sigma_1+\eta_2\sigma_2) }}
{\left(\cosh \pi (\sigma_1-\sigma_2+m)\cosh \pi (\sigma_1-\sigma_2-m)\right)^{N_f}}\ .
\ee
Introducing new integration variables as in (\ref{plusminus}), the partition function takes the form
\be
Z= 2 \int d\sigma_+ d\sigma_-  \ \frac{e^{-i\pi(k_1  +k_2  ) \sigma_-^2 -i\pi(k_1 +k_2) \sigma_+^2
-2 \pi i (k_1 -k_2)  \sigma_+\sigma_- +2\pi i (\eta_+ \sigma_+ +\eta_-\sigma_-) }}
{\left(\cosh \pi (2\sigma_- +m)\cosh \pi (2\sigma_- -m)\right)^{N_f}}\ ,
\label{uyt}
\ee
$$
\eta_- \equiv \eta_1-\eta_2\ ,\qquad \eta_+ \equiv \eta_1+\eta_2\ .
$$
We now  consider the specific model with parameters satisfying the relation
\be
k_1  +k_2 =0\ .
\label{solvc}
\ee
As a result, the $\sigma_+^2$ term in the exponent of (\ref{uyt}) cancels out and the integral over $\sigma_+$ gives a Dirac delta function.
If one considers chiral multiplets with generic gauge charges $(q_1,q_2)$ and  $(-q_1,-q_2)$ --thus maintaining the original normalization for the gauge fields-- the relation that eliminates the  $\sigma_+^2$ term from the exponent is
\be
\frac{q_1^2}{ k_1} +\frac{q_2^2 }{k_2} =0\ .
\label{solvcc}
\ee
Note that this condition requires that the Chern-Simons levels have opposite signs, i.e. $k_1 k_2<0$.

Returning to the condition (\ref{solvc}), this leads essentially to 
the mass-deformed ABJM case discussed earlier, where  it has now been extended to more flavors and to the case $\eta_1\neq \eta_2$.
The final expression for the partition function on the three-sphere is 
\be
Z=  \frac{1 }{|k_1|}
\ \ \frac{e^{i\pi  \frac{\eta_+ \eta_-}{k_1} }  }
{\left( \cosh \pi (  \frac{\eta_+}{k_1} + m) \cosh \pi (\frac{\eta_+}{k_1} - m)\right)^{N_f}}\ .
\ee

\subsubsection*{Partition function on the ellipsoid }

The calculation is similar, but now the basic building block in the one-loop determinant is the  double-sine function $s_b$. It is defined by 
\be
s_b(x) =\prod_{k,n=0}^\infty \frac{kb+nb^{-1} +Q/2-ix}{kb+nb^{-1} +Q/2+ix}\ ,\qquad Q=b+b^{-1}\ .
\ee
Then the one-loop determinant for a chiral field  of R-charge $\Delta $, gauge charges $(q_1,q_2)$ and mass $m$ is given by
\be
Z^{\rm chiral}_{{\rm 1-loop}} = s_b\left( \frac{iQ}{2}(1-\Delta )-q_1 \sigma_1-q_2\sigma_2+m\right)\ .
\ee
We consider $N_f$  chiral multiplets $\phi_r$ with  R-charge $\Delta $ and $U(1)\times U(1)$ gauge charges $(1,-1)$ and $N_a$  chiral multiplets $\tilde \phi_s$ with the same R-charge $\Delta $ and opposite gauge charges
$(-1,1)$.
In addition, with add mass deformation parameters $m_r,\ \tilde m_s$ satisfying $\sum_{r=1}^{N_f} m_i=0$,  $\sum_{s=1}^{N_a} \tilde m_s=0$.
Thus the total one-loop factor is given by
\be
Z^{\rm matter}_{{\rm 1-loop}} (\sigma_-,\Delta;m_i)= \prod_{r=1}^{N_f} 
s_b\left( \frac{iQ}{2}(1-\Delta )- 2\sigma_-+m_r\right)
\prod_{s=1}^{N_a} s_b\left( \frac{iQ}{2}(1-\Delta )+ 2\sigma_-+\tilde m_s\right)
\ee

For $k_1+k_2=0$, the partition function on the ellipsoid is  given by
\be
Z=2\int d\sigma_+ d\sigma_-     \ e^{-4 \pi i k_1 \sigma_+\sigma_- +2\pi i (\eta_+ \sigma_+ +\eta_-\sigma_-) }
Z^{\rm matter}_{{\rm 1-loop}} (\sigma_-,\Delta;m_i)\ ,
\ee
Thus
\be
Z=  \frac{1 }{|k|}
\ e^{i \pi  \frac{\eta_+ \eta_-}{k} }  
\prod_{r=1}^{N_f} 
s_b\left( \frac{iQ}{2}(1-\Delta )- \frac{\eta_+}{k}+m_r\right)
\prod_{s=1}^{N_a}s_b\left( \frac{iQ}{2}(1-\Delta )+ \frac{\eta_+}{k}+\tilde m_s\right)\ ,
\label{Zfin}
\ee
with $k\equiv k_1$.

The double-sine function can be written in another form, which is useful to study 
the limit $b\to 0$ (or, alternatively, $b\to \infty $), where  the ellipsoid degenerates to ${\mathbb R}^2\times {\mathbb S}^1$:
\be
s_b(x) = e^{- \frac{i\pi x^2}{2}}\prod_{k=1}^\infty  \left(1- w_1^{-(2k+1)} e^{-2\pi b x}\right)^{-1} \left(1- w_2^{-(2k+1)} e^{-2\pi  x/b}\right)^{-1}\ ,
\label{famor}
\ee
$$
w_1= e^{i\pi b^2}\ ,\qquad w_2= e^{i\pi /b^2}\ .
$$
In the present case, the partition function (\ref{Zfin})  contains contributions proportional to
\be
 e^{-2\pi b x}\sim e^{-\frac{2\pi b}{k}( \eta_+\pm k m_i)}   \ ,\qquad  e^{-2\pi  x/b}\sim e^{-\frac{2\pi}{bk}( \eta_+\pm k m_i)}
\ee
It would be interesting to understand the physical origin of these contributions.
As shown in the next section, these cannot be vortex contributions because the theory does not have supersymmetric vortices.
Indeed, the theory only admits the trivial vacuum with all
$\phi_r =\tilde \phi_s= 0$ (see section 5).

The key  point that allows one to explicitly carry out the two integrations in   (\ref{uyt})   is that, upon imposing  (\ref{solvc}), the integrand depends on one of the two integration variables, $\sigma_+$, 
only in the exponent, with linear dependence. The one-loop determinant does not depend on  $\sigma_+$, since the chiral matter
only couples to the vector multiplet $V_1-V_2$.
As discussed below, the underlying physical reason of the simplicity of these theories is that these are precisely the cases
where the theory does not have vortex configurations associated with north and south poles of the ellipsoid. 
It is worth noting that the simplicity of these theories is not related to possible enhancement of supersymmetries, that arises only for special matter content. In particular, if $N_f\neq N_a$, the theory always has ${\cal N}=2$ supersymmetry.

In more general 3d models
where (\ref{solvc}) is not satisfied, the partition function on the ellipsoid is given in terms of infinite sums where each term
represents a contribution from supersymmetric vortex configurations \cite{Pasquetti:2011fj}. 
This is evident in the ``Higgs branch localization" \cite{Fujitsuka:2013fga,Benini:2013yva}, where  another deformation term is added.
The localized field configuration is then  given in terms of vortex numbers. 
In the Coulomb branch localization, the equivalent result is obtained by computing the integrals by residue integration \cite{Pasquetti:2011fj,Chen:2013pha,Fujitsuka:2013fga,Benini:2013yva}.

As an example, we may consider the case where $q_2=0$.
In this case, one has $U(1)_{k_1}$ Chern-Simons-matter  plus a decoupled pure $U(1)_{k_2}$ CS sector without matter.
The non-trivial part in the partition function comes from the first sector. It is a particular case of the partition functions considered in
\cite{Benini:2013yva} for $U(N)$ CS theory coupled to $N_f$ fundamentals and $N_a$ antifundamentals.
In our case, $N_f$, $N_a$ correspond to the number of chiral multiplets with charge $q_1$ and $-q_1$, respectively.
The partition function is then given by particularizing (2.75) of \cite{Benini:2013yva} to $N=1$.
One obtains an expression for $Z$ as  a product of vortex and antivortex partition functions $Z_{\rm v},\ Z_{\rm av}$. In particular, in the  case with only FI mass deformation, one finds an expression of the form 
\be
Z_{\rm v}=\sum_{n=0}^\infty e^{-2\pi b^{-1}\eta_1 n}z_{\rm v}^{(n)}(b)\ ,\qquad Z_{\rm av}=\sum_{n=0}^\infty e^{-2\pi b\eta_1 n}z_{\rm av}^{(n)}(b)\ ,
\ee
where $n$ is identified with the absolute value of the vortex topological charge. We see that 
the 
vortex and antivortex actions have the expected linear dependence with the FI parameter and linear dependence with the topological charges 
\cite{Pasquetti:2011fj}.

\section{Flat space analysis}

Here we will show that vortex configurations disappear {\it precisely} in the case
when the partition function reduces to a single term due to the condition (\ref{solvc}),
\be
k_1  +k_2  =0\ .
\nonumber
\ee
In flat spacetime, the FI term is
\be
S_{\rm FI}[\eta_a]= -i  \int d^3 x   (\eta_1D_1+\eta_2 D_2)  \ .
\ee
The part of the action involving $D_1,\ D_2,\ \sigma_1,\ \sigma_2$ is
\bea
S' &=&   \int d^3 x   
\Big(-2ik_1 D_1\sigma_1-2ik_2 D_2\sigma_2- i(\eta_1 D_1+\eta_2 D_2)
\nonumber\\
&+& i(D_1- D_2) \bar\phi \phi +(\sigma_1-\sigma_2)^2\bar\phi\phi\Big) \ .
\eea
The equations for $D_1,\ D_2$ give
\be
\label{solsig}
-2k_1 \sigma_1-\eta_1 +\bar\phi \phi=0\ ,\qquad -2k_2 \sigma_2 -\eta_2 -\bar\phi \phi=0\ .
\ee
It follows that $2k_1\sigma_1+2k_2\sigma_2=-\eta_1-\eta_2 =-\eta_+={\rm const.}$

The equations of motion for $\sigma_1, \ \sigma_2$ give
\be
\label{solD}
ik_1 D_1-\bar\phi \phi (\sigma_1-\sigma_2)=0\ ,\qquad ik_2 D_2 
+\bar\phi \phi (\sigma_1-\sigma_2)=0\ .
\ee
Therefore, $k_1  D_1=-k_2  D_2$.
When there are several copies of scalar fields $\phi_r $, $r=1,...,N_f$, with charges $(1,-1)$, and $\tilde \phi_s $, $s=1,...,N_a$, with charges $(-1,1)$,
then the above equations generalize as follows
\bea
&& -2k_1 \sigma_1-\eta_1 +\sum_r |\phi_r |^2- \sum_s |\tilde\phi_s |^2=0\ ,
\nonumber\\
&& -2k_2 \sigma_2 -\eta_2 
-\sum_r |\phi_r |^2+ \sum_s |\tilde\phi_s |^2 =0\ ,
\nonumber\\
&& ik_1 D_1- \Big( \sum_r |\phi_r |^2+ \sum_s |\tilde\phi_s |^2\Big) (\sigma_1-\sigma_2)=0\ ,
\nonumber\\
&& ik_2 D_2 
+ \Big( \sum_r |\phi_r |^2+ \sum_s |\tilde\phi_s |^2\Big) (\sigma_1-\sigma_2)=0\ .
\nonumber
\eea

We look for supersymmetric configurations. In the flat limit, the supersymmetric transformations (\ref{ate}), (\ref{bate}) become
\bea
&&\delta \lambda_1 =\left( \frac{1}2 \epsilon_{\mu\nu\rho}F^{\nu\rho}[A_1] -\p_\mu \sigma_1
\right) \gamma^\mu \epsilon - iD_1\epsilon\ ,
\nonumber\\
&&\delta \lambda_2 =\left( \frac{1}2 \epsilon_{\mu\nu\rho}F^{\nu\rho}[A_2] -\p_\mu \sigma_2
\right) \gamma^\mu \epsilon - iD_2\epsilon\ ,
\nonumber\\
&&\delta \psi = -\gamma^\mu \epsilon D_\mu \phi -\epsilon(\sigma_1-\sigma_2)\phi +i\bar \epsilon F\ .
\eea
Recall $D_\mu\phi = \left( \partial_\mu -i  (A_1)_\mu + i (A_2)_\mu \right) \phi $. 
We must impose $\delta \lambda_1 =\delta \lambda_2=\delta \psi_r=\delta \tilde \psi_s=0$.
Considering the equation $k_1 \delta \lambda_1+ k_2 \delta \lambda_2=0$, we deduce that
\be
k_1 F^{\nu\rho}[A_1]=-k_2  F^{\nu\rho}[A_2]\ ,
\ee
i.e.
\be
F^{\nu\rho}[\tilde A]\equiv 0\ ,\qquad \tilde A \equiv k_1 A_1+k_2  A_2\ .
\ee
The scalar field $\phi $ couples to the gauge field
$$
\hat A= A_1- A_2\ .
$$
A vortex solution $\phi = f(r) e^{in \varphi}$ implies a circulation for $\hat A$ and, by Stokes theorem,  a flux $F_{12}[\hat A]\propto n\neq 0$.
However, this is impossible if $\hat A$ is proportional to $\tilde A$, since $F_{12}[ \tilde A]= 0$.
These gauge fields are proportional to each other when 
\be
k_1  +k_2  =0\ ,
\nonumber
\ee
which is nothing but the same condition (\ref{solvc}) that leads to a simple partition function with a single term.
In the next subsection we will re-derive this condition from the effective potential.

The resulting theory with $k_1+k_2=0$ can be cast in a familiar form.
Introducing new vector multiplets $V_A=V_1-V_2\equiv (A_\mu,\ \lambda_A,\ \bar\lambda_A,\ \sigma_A,\ D_A)$ and 
$V_B=V_1+V_2\equiv (B_\mu,\ \lambda_B,\ \bar\lambda_B,\ \sigma_B,\ D_B)$
the action becomes
\be
\label{BFact}
S=ik \int B\wedge dA-ik \int d^3 x\sqrt{g}\left(-\bar\lambda_B \lambda_A-\bar\lambda_A\lambda_B+D_B\sigma_A+D_A\sigma_B\right)
+S_{\rm matter}[V_A]\ ,
\ee
where the matter action is given by (\ref{mmmm}) by replacing $\hat V $ by $V_A$.
This is nothing but a BF Chern-Simons model with matter coupled to  only one of the two gauge fields.
One can  directly see that there are no supersymmetric vortex solutions. The equations of motion of $B_\mu $ set
\be
F_{\mu\nu}[A]=0\ .
\label{ffgg}
\ee
Considering now the supersymmetric variation $\delta\psi =0$, one finds that preserving 1/2 of the supersymmetries requires
\be\label{casr}
(D_1+iD_2)\phi =0 \qquad {\rm or} \qquad (D_1-iD_2)\phi =0\ .
\ee
In either case, by (\ref{casr}),
a solution with non-trivial topological phase, $\phi=f(r)e^{in\varphi}$, implies 
$\oint dx^i A_i=2\pi n$, in contradiction with (\ref{ffgg}). Thus the topological charge must be zero.

\subsubsection*{Standard vortex solutions in the general case}

Let us consider the general model with arbitrary Chern-Simons levels $k_1,\ k_2$.
From (\ref{solsig}), (\ref{solD}) one can express $D_1,\ D_2, \sigma_1,\ \sigma_2$ in terms of $\bar\phi\phi $.
Substituting the solution into the action, the bosonic part of the Euclidean action takes the form
\be
S_{\rm bos}= i k_1\int A_1\wedge A_1 + i k_2\int A_2\wedge A_2
+ \int d^3x \left( D_\mu[\hat A] \bar\phi D_\mu[\hat A]\phi +V_{\rm eff}(\phi)\right)\ ,
\ee
$$
D_\mu[\hat A]\phi =
\partial_\mu \phi -i \hat A_\mu\phi \ ,\qquad \hat A= A_1- A_2\ ,
$$
where 
\be
V_{\rm eff}=  \frac{1}{4k_1^2k_2^2} \bar\phi \phi\  \left( -\eta_0 +  (k_1+k_2)\ \bar \phi \phi \right)^2 \ ,
\ee
\be
\eta_0 \equiv  k_2 \eta_1- k_1 \eta_2 \ .
\ee
When there are several copies of  scalar fields $\phi_r, \ \tilde \phi_s   $ with charges $(1,-1)$ and $(-1,1)$,
$r=1,...,N_f$, $s=1,...,N_a$, the potential becomes
\be
V_{\rm eff}=  \frac{1}{4k_1^2k_2^2} \left( | \phi_r|^2 + |\tilde \phi_s |^2 \right)\  \left(
-\eta_0 +  (k_1+k_2)\  \left( | \phi_r|^2 - |\tilde \phi_s |^2 \right) \right)^2 \ ,
\ee
where sums over $r$ and $s$ are understood
(i.e. $|\phi_r|^2 =\sum_{r=1}^{N_f} \bar\phi_r \phi_r$, $|\tilde \phi_s|^2 =\sum_{s=1}^{N_a} \bar{\tilde \phi}_s \tilde \phi_s$).

We now see the physical origin of the absence of vortices when $k_1+k_2=0$: the potential becomes 
$$
V_{\rm eff}\to \frac{\eta_0^2}{4k_1^2k_2^2}  \left( | \phi_r|^2 + |\tilde \phi_s |^2 \right)\ .
$$ 
This potential has only the trivial vacuum $\phi_r =\tilde\phi_s= 0$.
In the particular case of ABJM theory \cite{abjm}, this just reflects the familiar feature
 that in the abelian $U(1)\times U(1)$ case the sixth-order potential vanishes,
leaving only the mass deformations.\footnote{Studies of vortex configurations in non-abelian $U(N)\times U(N)$ ABJM theory can be
found in \cite{Kim:2009ny,Mohammed:2012rd}.}

Let us introduce the gauge field
$$
B=  A_1+ A_2 \ .
$$
The part of the action containing the vector bosons takes the form
\bea
S &=& \frac{i}{4}\int \left( (k_1 +k_2 ) \hat A \wedge d\hat A +(k_1 +k_2 )  B\wedge dB 
+2(k_1-k_2)  B \wedge d\hat A \right)
\nonumber\\
&+&\int d^3x  D_\mu[\hat A ] \bar\phi D_\mu[\hat  A]\phi\ .
\eea
Now $B_\mu $ can be integrated out by its equation of motion. One obtains
\be
(k_1 +k_2 ) dB +(k_1-k_2)  d\hat A=0\ .
\ee
In the special case when $k_1 +k_2 =0$, we recover the condition found above that the flux $d\hat A =0$. This is the theory with no vortex
configurations. For $k_1 +k_2 \neq 0$, we can solve the above equation for $B_\mu $ and find the bosonic (Euclidean) action
\be
S=i\tilde k\int \hat  A \wedge d\hat A +
\int d^3x \left( D_\mu[\hat A] \bar\phi D_\mu[\hat A]\phi+  \frac{1}{4\tilde k^2} \bar\phi \phi\  \left( \bar \phi \phi - \frac{\eta_0}{  k_1+k_2} \right)^2\right)\ ,
\ee
where
\be
\tilde k =\frac{k_1 k_2}{(k_1 +k_2 )}\ , \qquad k_1+k_2 \neq 0\ .
\ee
We recognize the  Chern-Simons-Higgs action for a $U(1)$ gauge group with a sixth-order Higgs potential. 
This theory has been studied extensively in the literature \cite{Hong:1990yh,Jackiw:1990aw}. 
Indeed, the potential is the same effective potential that arises from ${\cal N}=2$ supersymmetric $U(1)$ CS theory coupled to chiral matter --with the precise overall coefficient required by supersymmetry \cite{Lee:1990it}
(for recent discussions, see \cite{Kim:2012uz,Arias:2015fsa} and references therein).
The theory has well-known vortex configurations due to the existence of a  non-trivial $U(1)$ symmetry-breaking vacuum $\bar\phi\phi =\eta_0/(k_1+k_2)$, provided $\eta_0/(k_1+k_2)>0$. More generally, in the presence of scalar fields $\phi_r,\ \tilde\phi_s$, there are $U(1)$ symmetry-breaking vacua
for any sign of   $\eta_0/(k_1+k_2)$,  satisfying $ \langle \sum_r | \phi_r|^2 -\sum_s |\tilde \phi_s |^2 \rangle=\eta_0/(k_1+k_2)$.

On ${\mathbb R}^2\times {\mathbb S}^1_{\beta}$, the  Euclidean action for a vortex configuration of vortex number $n$ is given by \cite{Hong:1990yh,Jackiw:1990aw}
\be
S= 2\pi \beta \frac{\eta_0}{k_1+k_2}\ n\ .
\label{sacfin}
\ee
From a more physical perspective, one can see  why vortices are forbidden in the theory with $k_1+k_2=0$.
In the limit $k_1+k_2\to 0$, the action of a vortex is infinity.
The result (\ref{sacfin}) may be compared with the vortex action obtained from the ellipsoid partition function of the $U(1)$ CS model discussed above, $S=2\pi \eta_1 n/b $ or
$S=2\pi b \eta_1 n $ for $b\to 0,\infty $. The effective potentials are the same, with the parameter $\eta_1 $ identified with $\eta_0/(k_1+k_2)$.\footnote{Another way to see this is by restoring the dependence on the original gauge charges $q_1,q_2$, by rescaling $k_1\to k_1/q_1^2$,  $k_2\to k_2/q_2^2$, $\eta_1\to \eta_1/q_1$, $\eta_2\to -\eta_2/q_2$.
Then $S= 2\pi \beta \frac{k_2 q_1 \eta_1+k_1 q_2\eta_2}{k_1q_2^2+k_2q_1^2}\ n$.  The $U(1)$ CS-matter theory is
then obtained for $(q_1,q_2)=(1,0)$, 
again giving 
 $S= 2\pi \beta \eta_1 n$.}
Thus the vortex actions agree with the identification $\beta \to b$ or $\beta \to 1/b$ (this is of course the observation in \cite{Pasquetti:2011fj,Fujitsuka:2013fga,Benini:2013yva}, now adapted to our context). 

\medskip

In conclusion, the theory with $k_1+k_2\neq 0$ is essentially equivalent to a $U(1)$ Chern-Simons-matter theory plus a decoupled
$U(1)$ pure CS sector. This theory has vortices. The theory with $k_1+k_2=0$ is special: it does not have vortices, nonetheless it has a 
non-trivial partition function (\ref{Zfin}),  containing non-perturbative contributions in the FI coupling.
It would be very interesting to clarify the origin of such contributions and to have a physical understanding
of the structure of  (\ref{Zfin}). 

\bigskip

Finally, let us consider CS-matter theories with gauge group $U_{k_1}(1)\times \cdots \times U_{k_N}(1)$ and $N_f$ chiral multiplets with the
same charges $(q_1,...,q_N)$.
Then a straightforward generalization of the above discussion gives the potential
\be
V_{\rm eff}=  \frac{1}{4} \left( \sum_{r=1}^{N_f} | \phi_r|^2 \right)\  \left(
-\eta_0 +  c\  \sum_{r=1}^{N_f} | \phi_r|^2  \right)^2 \ ,
\label{poteu}
\ee
where
\be
\eta_0 \equiv \sum_{a=1}^N \frac{q_a \eta_a}{k_a}\ ,
\qquad c\equiv  \sum_{a=1}^N \frac{q_a ^2}{k_a}\ .
\ee
It follows that the potential simplifies when
$c=0$. In this case the potential becomes quadratic and the theory does not have vortices.
In particular, if all charges $q_a$ are different from zero we can normalize the vector fields by setting $q_a =\pm 1$.  
Then the no-vortex condition becomes
\be
\sum_{a=1}^N \frac{1}{k_a} = 0\ .
\label{reop}
\ee
Clearly, the same properties hold if supersymmetric mass deformations are added 
(contributing as $m_r^2\bar \phi_r\phi_r$ to the bosonic potential (\ref{poteu})).

Like in the $U(1)\times U(1)$ case, when $c=0$ the partition function $Z$ dramatically simplifies. The partition function on ${\mathbb S}^3_b$ is given by
\be
Z=\int d^N\sigma \ e^{-i\pi \sum_a k_a \sigma_a^2
+2\pi i \sum_a \eta_a \sigma_a}\ Z_{\rm 1-loop}\big(\hat\sigma;\ \Delta,\ b,\ m_r\big) \ ,
\ee
$$
\hat \sigma \equiv \sum_{a=1}^N q_a\sigma_a\ .
$$
Since the one-loop determinant depends on $\sigma_a$ only through the combination $\hat \sigma $, it is convenient to introduce new integration variables $\sigma_1, ...,\sigma_{N-1},\hat \sigma$. Then the integrations over $\vec\sigma\equiv (\sigma_1, ...,\sigma_{N-1})$ only involve a Gaussian factor $\exp(\vec \sigma^{\rm T}.M.\vec\sigma + \vec V.\vec\sigma )$,
where $M$ is an $(N-1)\times (N-1)$ matrix.
However,  when $c=0$, the determinant of $M$ vanishes. Therefore $M$ has (at least) one zero eigenvalue.
As a result, one can perform $N-2$ Gaussian integrations and the remaining integration
over the eigenvector $\tilde \sigma $ with vanishing eigenvalue yields a delta function, of the type $\delta \big(\hat \sigma-\hat\sigma_0(\eta_a,k_a,q_a)\big)$.
Thus the complete integral can be carried out explicitly, just as in the $U(1)\times U(1)$ case, giving rise to a compact expression.
For example, setting all $q_a =1$, under the condition (\ref{reop}), one finds a delta function setting 
$$
\hat \sigma\to \hat\sigma_0 \equiv \sum_{a=1}^N \frac{\eta_a}{k_a}\ .
$$
The final result is
\be
Z=\frac{e^{i\pi \sum_a \frac{\eta_a^2}{k_a}}} {\sqrt{ \prod_{a=1}^N |k_a|}}\ 
Z_{\rm 1-loop}\big(\hat \sigma_0;\ \Delta,\ b,\ m_r\big)\ .
\ee
 One can also extend the model by adding $N_a$ chiral multiplets with opposite charges $(-q_1,...,-q_N)$ with similar results.


\section*{Acknowledgements}

We are grateful to M. Honda for useful
discussions.
J.G.R. would like to thank the
Department of Physics, FCEN of Universidad de Buenos Aires, for hospitality during the
course of this work. He also acknowledges financial support from projects  FPA2013-46570 (MINECO), 2014-SGR-104 (Generalitat de Catalunya) 
and  MDM-2014-0369 of ICCUB (Unidad de Excelencia `Mar\'ia de Maeztu'). F.A.S.  acknowledges financial support from ANPCyT, CICBA, CONICET.

\end{document}